\documentstyle[12pt,epsf]{ioplppt}

\newcommand \sign           {\mathop{\rm sign}\nolimits}

\newcommand \nn             {\nonumber}

\newcommand{\be}{\begin{equation}}
\newcommand{\ee}{\end{equation}}
\newcommand{\bea}{\begin{eqnarray}}
\newcommand{\eea}{\end{eqnarray}}
\newcommand{\xiimu}{\xi_i^{\mu}}

\newcommand	{\klm}[1]     {{ \left( {#1} \right) }}
\newcommand	{\ekl}[1]     {{ \left[ {#1} \right] }}
\newcommand	{\gkl}[1]     {{ \left\{{#1} \right\}}}
\newcommand	{\skl}[1]     {{ \left\langle{#1} \right\rangle}}
\newcommand	{\btr}[1]     {{ \left|{#1} \right|}}

\begin{document}
\jl{1}
\letter{Storage of correlated patterns in a perceptron}
\author{B L\'opez, M Schr\"oder and M Opper}
\address{Institut f\"ur Theoretische Physik, Universit\"at W\"urzburg,
Am Hubland, \\ D-97074~W\"urzburg}
\begin{abstract}
We calculate the storage capacity of a perceptron for correlated
gaussian patterns. We find that the storage capacity $\alpha_c$ can
be less than 2 if similar patterns are mapped onto different outputs
and vice versa. As long as the patterns are in general position 
we obtain,
in contrast to previous works, that $\alpha_c \geq 1$ in agreement
with Cover's theorem. Numerical simulations confirm the results.
\end{abstract}

\pacs{07.05.Mh, 05.20.-y, 05.90.+m, 87.10.+e}


\medskip

The critical storage capacity of a simple perceptron for randomly
chosen input/output pairs is known to be $\alpha_{c} = p/N = 2$, with
$p$ the number of stored patterns and $N$ their input dimension. This result
was first derived by Cover (1965) using a geometrical argument and
later by Gardner (1988) and Gardner and Derrida (1988) by calculating the 
fractional phase space volume of consistent couplings with the tools of 
statistical mechanics and the replica approach.

Cover's theorem states that, as long as the patterns $\vec{\xi}^{\mu}
(\mu = 1, \ldots, p)$ are in general position (no subset of $N$ or less
patterns is linear dependent), the critical storage capacity $\alpha_{c}$
is at least 1, independently of the corresponding outputs $s^{\mu}$.
For $\alpha$ between 1 and 2 the fraction of output combinations which
are not linearly separable is exponentially small in $N$. The converse
holds for $\alpha$ larger than 2. In this case a randomly chosen sequence 
of outputs will not be linearly separable with a probability approaching
1 as $N \rightarrow \infty$. This is even true if correlations among the 
patterns $\vec{\xi}^{\mu}$ are introduced (Monasson 1992).

One would argue that in general correlations which include the outputs lead to
higher critical storage capacities as it is for instance the case for
biased patterns (Gardner 1988).
Recently it has been found (Bork 1994, Schr\"oder \etal 1995) that patterns 
and outputs 
extracted from a bit-sequence seem to lead to smaller storage capacities than
$\alpha_{c} =2$ for a perceptron. The bit-sequence is thought as an infinite
time series $\zeta_i \ (i=1,2,\ldots)$ in which the first pattern results from
the first $N$ values $\zeta_i \ (i=1,\ldots,N)$ and its output from the 
$(N+1)$th value  $\zeta_{N+1}$ (or $\sign(\zeta_{N+1})$ for continuous valued 
$\zeta_i$). By moving this $(N+1)$-broad window one step forward, the second
pattern and its corresponding output result, and so on. If the $\zeta_i$
are drawn at random from a distribution with $\skl{\zeta_i}=0,
\skl{\zeta_i^2}=1$, one finds that the critical storage capacity of a perceptron 
which stores the resulting input/output pairs $(\vec{\xi}^{\mu},s^{\mu})$
is   $\alpha_{c} \approx 1.82$ for binary $\zeta_i$ and 
$\alpha_{c} \approx 1.88$ for gaussian $\zeta_i$. 
This result indicates that
the embedded correlations between input/output pairs resulting from the
bit-sequence lead to a reduction of the storage capacity when compared
to randomly chosen pairs and are harder to implement in a perceptron.

Taking the former as a motivation, we will investigate in the present letter
the effect of correlations between input/output pairs on the critical storage
capacity of a perceptron.
The main idea is that the storage of similar patterns with different outputs
should be more difficult to implement in a perceptron than the case of
similar patterns with identical outputs. 
If we introduce the transformed patterns 
$\vec{\sigma}_{\mu} = \vec{\xi}_{\mu} s^{\mu}$,
this similarity or dissimilarity can be described with a positive or
negative overlap 
($R = N^{-1} \vec{\sigma}_{\mu} \cdot \vec{\sigma}_{\nu}$) respectively
between two transformed
patterns $\mu$ and $\nu$.  Without loss of generality 
we fix all outputs to have the value $s^{\mu}= +1 \ (\mu = 1, \ldots, p)$, so
that $\vec{\sigma}_{\mu} = \vec{\xi}_{\mu}$.
Let us now take pairs of patterns with a fixed overlap $R$. With a normalization
such that  $\vec{\xi}_{\mu}^{\ 2} = N$, we have:
\be
\label{defcorr}
\vec{\xi}_{2 \mu -1} \cdot \vec{\xi}_{2 \mu} = N R  \quad  
\rm{for} \quad  \mu=1,\ldots,p/2
\ee
with $\btr{R} \leq 1$. Apart from these fixed overlaps two arbitrary patterns 
shall not be correlated and thus will have an overlap of 
$N^{-1} \vec{\xi}_{\mu} \cdot \vec{\xi}_{\nu} = 0$ in the thermodynamic limit
$N \rightarrow \infty$. 

Before evaluating the storage capacity for a general $R$ let us consider
some special cases of interest. For $R=1$ the two patterns out of a pair
are identical (for $N \rightarrow \infty$), so that the storage of the first
pattern automatically implements the second one, concluding that 
$\alpha_{c}\klm{R=1} =4$. For $R=0$ no correlations are present and
$\alpha_{c}\klm{R=0} =2$ as usual. For $R=-1$ the patterns are not in general
position, since $p/2$ of them are pairwise linearly dependent. Already the 
first two patterns cannot be implemented by a perceptron, so that
$\alpha_{c}\klm{R=-1} =0$. However, if $R$ is very close to $-1$ general position
is guaranteed and thus $\alpha_{c}\klm{R=-1+\epsilon} \geq 1 ,\ 
\klm{\epsilon > 0}$ 
according to Cover's theorem. We will investigate this case later in more detail. 

Following Gardners approach we consider the fractional phase space volume
$V$ of couplings $\vec{J}$ that are consistent with the constraints imposed by
the patterns (version space):
\be
\label{verspace}
V = V_{tot}^{-1}\ \int \prod_{i=1}^N dJ_i\
                  \prod_{\mu = 1}^p \Theta \left (
                   s^{\mu} \frac1{\sqrt{N}}\ \sum_{i=1}^N J_i \xiimu - \kappa
                                           \right )\ 
                  \delta \left (
                     \sum_{i=1}^N J_i^2  -   N 
                         \right ) \ ,       
\ee
with
\be
\label{vtot}
V_{tot} = \int \prod_{i=1}^N dJ_i\
               \delta \left (
                     \sum_{i=1}^N J_i^2  -   N 
                      \right ) \ .  
\ee
It is clear, that for negative $R$ the reduction of the version space by every 
new pair of patterns is more drastic than for $R=0$ and so we expect 
$\alpha_{c}$ to be less than 2.

We now fix  $s^{\mu}= +1 \ (\mu = 1, \ldots, p)$ and
introduce the conditions \eref{defcorr} via delta functions. 
The average $\skl{\skl{ \mbox{ln} V }}_{\xi}$ will be performed by means of
the replica trick and is defined by:
\bea
\label{average}
\fl
\skl{\skl{ f \klm{ \gkl{\vec{\xi}^{\nu}}_{\nu =1}^p } }}_{\xi} = \nn\\
        C_N \int \prod_{\mu = 1}^{p/2} \prod_{i=1}^N 
        \klm{ D \xi_i^{2 \mu -1} \ D \xi_i^{2 \mu}} \
        f\klm{ \gkl{\vec{\xi}^{\nu}}_{\nu =1}^p } \ 
        \prod_{\mu = 1}^{p/2} \delta 
        \klm{ N^{-1} \vec{\xi}_{2 \mu -1} \cdot \vec{\xi}_{2 \mu} - R }
\eea
where $C_N$ is the normalization constant resulting from 
$\skl{\skl{ 1 }}_{\xi} = 1$, and 
$Dx \equiv dx \exp \klm { -x^2/2}/ \sqrt{2 \pi}$ is the Gaussian measure.
As usual in the calculation the order parameter $q_{\alpha \beta}$ appears,
which measures the overlap between two replicas. 
Making the replica symmetric ansatz $q_{\alpha \beta} = q \ (\forall 
\alpha < \beta)$ and using 
$\skl{\skl{ \rm{ln} V }}_{\xi} = 
 \lim_{n \rightarrow 0} n^{-1} {\rm ln}\skl{\skl{ V^n }}_{\xi} $
together with the saddle point method to evaluate the integrals in the
thermodynamic limit, one obtains:
\be
\label{entropy}
\fl
\frac1N \skl{\skl{ \rm{ln} V }}_{\xi} = 
\mathop{{\rm Extr}}_{q} \gkl{ \frac{\alpha}2 \int Dx \int Dz \
                {\rm ln} f \klm{q,R,\kappa,x,z}
                + \frac12\ {\rm ln} \klm{1-q}
                + \frac12 \frac{q}{1-q}         }
\ee
with
\be
f \klm{q,R,\kappa,x,z} = \sqrt{1-R^2} \int_{\gamma}^{\infty} D\omega
                                      \int_{\delta}^{\infty} Du \
                                      \exp \klm{R u w}
\ee
where
\be
\gamma = \frac{\kappa + \sqrt{q}\ x}
              {\sqrt{1-q}\ \sqrt{1-R^2} } \quad , \quad
\delta = \frac{\kappa + \sqrt{q}\ \klm{R x + \sqrt{1-R^2}\ z}}
              {\sqrt{1-q}\ \sqrt{1-R^2}}
\ee
The critical storage capacity is reached when the version space shrinks to a 
single point and thus $q$ reaches unity. From the extremum condition in 
\eref{entropy} we obtain for the critical storage capacity $\alpha_c$
\be
\alpha_c^{-1} \klm{R,\kappa} = \lim_{q \rightarrow 1}
\ekl{ - \klm{1-q}^2 \int Dx \int Dz \ \frac{\partial}{\partial q}
{\rm ln} f\klm{q,R,\kappa,x,z}  }                                    
\ee
In the limit $q \rightarrow 1$ one has
to consider the cases of positive and negative $R$ separately yielding
\bea
\fl
\alpha_c^{-1} \klm{R,\kappa} = 
\int_{- \infty}^{x_0} Dx \int_{z_0}^{\infty} Dz \ \frac{b^2}2  \ + \
\int_{x_0}^{\infty} Dx  \int_{- \infty}^{z_1} Dz \ \frac{a^2}2   
\nonumber \\
+\ \int_{x_0}^{\infty} Dx  \int_{z_1}^{z_2} Dz \ 
                           \frac{a^2-2Rab+b^2}{2\ \klm{1-R^2}} \ + \
\int_{x_0}^{\infty} Dx  \int_{z_2}^{\infty} Dz \ \frac{b^2}2 
  \qquad  & \klm{R>0} 
\nonumber \\
\fl
\alpha_c^{-1} \klm{R,\kappa} =
\int_{- \infty}^{x_0} Dx \int_{z_0}^{z_2} Dz \ \frac{b^2}2  \ + \
\int_{- \infty}^{x_0} Dx \int_{z_2}^{\infty} Dz \ 
                              \frac{a^2-2Rab+b^2}{2\ \klm{1-R^2}}
\nonumber \\
+\ \int_{x_0}^{\infty} Dx  \int_{- \infty}^{z_1} Dz \ \frac{a^2}2 \ + \
\int_{x_0}^{\infty} Dx  \int_{z_1}^{\infty} Dz \ 
                               \frac{a^2-2Rab+b^2}{2\ \klm{1-R^2}}
\qquad  & \klm{R<0}
\nonumber 
\eea
with $a=\kappa + x$ , $b=\kappa + R\, x + \sqrt{1-R^2}\, z$ and
\be
\fl
z_0 = - \frac{\kappa + R\, x}{\sqrt{1-R^2}} \ , \
z_1 = - \frac{\kappa\, \klm{1-R}}{\sqrt{1-R^2}} \ , \
z_2 = \frac{\kappa\, \klm{1-R} + x\, \klm{1-R^2}}{R\, \sqrt{1-R^2}}
\ee
For the interesting case $\kappa=0$ this expression reduces to a surprisingly
simple form:
\be
\label{alphac}
\alpha_c^{-1} \klm{R,\kappa=0} \ =\  \klm{\frac14 + \frac{\phi}{2 \pi}} \qquad , \
\qquad \phi = \arccos R
\ee
Here $\phi$ is the angle between two correlated patterns, and \eref{alphac}
holds for $-1 < R \leq 1$. The resulting curve is plotted in \fref{alpha2}.
The simulations have been carried out in two ways: First, one can calculate
the average probability, that a given set of patterns with correlations 
as described above, is linearly separable for different values of $\alpha$.
The condition for the critical capacity is that this probability equals $1/2$.
The second method is to assume that the median learning time (for the
perceptron learning rule) scales as $\tau^{-0.5}\sim(\alpha_c-\alpha)$ for
$\alpha\to\alpha_c$. This can be proven for uncorrelated patterns (Opper 1988)
and has been used by Priel \etal (1994) to determine the critical capacity
by extrapolating the curve to $\tau\to\infty$. The inset of \fref{alpha2}
shows clearly that the mentioned scaling law is obeyed in our case too.

As we expected, for negative $R$ the storage capacity lies below 2, approaching
the value of $\alpha_c=4/3$ for $R \rightarrow -1$. 
This result can be understood as follows. Every pair $\mu = 1, \ldots, p/2$
of correlated patterns defines a vector $\vec{L}_{\mu} = \vec{\xi}_{2 \mu -1} - 
\vec{\xi}_{2 \mu}$. For $R \rightarrow -1$ the coupling vector $\vec{J}$ falls
for every $\mu$ into a hyperhalfplane orthogonal to $\vec{L}_{\mu}$. This leads
to the constraints:
\bea
\label{geometric}
(1) \quad \vec{L}_{\mu} \cdot \vec{J} \rightarrow &0   \nonumber \\
(2) \quad \vec{T}_{\mu} \cdot \vec{J} \geq &0 \qquad \mu = 1, \ldots, p/2
\eea
for $\vec{J}$, where $\vec{T}_{\mu}$ is a vector orthogonal to $\vec{L}_{\mu}$.
The second constraint defines a new perceptron problem with $p/2$ uncorrelated
patterns $\vec{T}_{\mu}$ for a coupling vector $\vec{J}$ with an effective
number of dimensions of $\klm{N-p/2}$ due to the first constraint. According to
Cover's theorem $\klm{p/2}/\klm{N-p/2} = 2$ and thus $\alpha_c=p/N=4/3$.

Up to now we have assumed that the overlap $R$ is the same for all pairs.
The case where two patterns have
an overlap $R$ with a probability $P\klm{R}$ can also be handeled.
The averages over different
values of $R$ factorize and the storage capacity is simply given by
\be
\label{alphadist}
\fl
\alpha_c^{-1} \klm{\kappa =0}  =  
                 \int_{-\infty}^{\infty} dR\ P\klm{R}\ 
                      \klm{\frac14 + \frac{\phi}{2 \pi}} 
            =   \int_{-\infty}^{\infty} dR\ P\klm{R}\  
                   \alpha_c^{-1} \klm{R,\kappa=0}  
\ee
The last equality states that the reciprocal values of the capacity for
a fixed value of the overlap weighted with their corresponding probability
sum up to the reciprocal of the total capacity.
If the distribution is symmetric $(P\klm{R} = P\klm{-R})$ the storage capacity
is $\alpha_c=2$. This is immediately clear if one thinks of a correlation
with a primary distribution $P_1\klm{R}$ and in addition chooses outputs $s^{\mu}$
at random $\pm 1$ instead of taking them all equal to $+1$. The new distribution
of the correlations is then given by $P\klm{R} = P_1\klm{R} / 2 
+ P_1\klm{-R} / 2$, which is symmetric and will lead in \eref{alphadist} to  
$\alpha_c=2$ as it should, since we have taken the outputs to be random. 

Let us now turn to the case where more than two patterns are correlated
among each other. In the case of three patterns with equal pairwise overlap
$R$ we can conclude that for $R=1$ the storage capacity is
$\alpha_c=6$ for $\kappa = 0$. If $R$ tends to $-1/2$ which is the minimal
accessible value in this case, a geometrical argument similar to \eref{geometric} leads
us to $\klm{p/3}/\klm{N-2p/3} = 2$ and thus $\alpha_c=p/N=6/5$.
The calculation of $\alpha_c\klm{R}$ for other values of $R$ should be more 
complicated than in the
the former case, since to disentangle the additional correlations one has to
introduce more gaussian fields. 
In general, if the patterns are correlated in tupels of $m$ it follows in the
same way:
\bea
\label{specialR}
\alpha_c\, \klm{R=1,\kappa=0} &=& 2 m    \nonumber \\
\alpha_c\, \klm{R=0,\kappa=0} &=& 2      \nonumber \\
\alpha_c\, \klm{R \rightarrow \frac{-1}{m-1},\kappa=0} &=& \frac{2m}{2m-1}
\eea
For $m \rightarrow \infty$, $\alpha_c$ tends to 1 as $R$ approaches the
minimal possible value $-1/(m-1)$.

As long as $m$ is of the order 1 compared to the number of patterns $p$, we
will have to calculate $\alpha_c\, \klm{R}$ as for $m=2$, which for larger
$m$ becomes very complicated. If $m$ however is of the order $p^{\gamma}$
with $0<\gamma \leq 1$ we are able to proceed in a much simpler way. For
this purpose we first introduce as in (Fontanari and Meir 1989) the 
correlation matrix $C$ for the patterns:
\be
c_{\mu \nu} = \skl{ \xi_i^{\mu} \xi_i^{\nu} }_{\xi} \qquad
\forall \mu, \nu = 1,\ldots ,p
\ee
The $\vec{\xi}_i = \klm{\xi_i^1, \ldots , \xi_i^p}$ are then distributed as
\be
\label{avergauss}
P\klm{\vec{\xi}_i} = \frac1{\sqrt{\klm{2 \pi}^p \det C}} \ 
                     \exp \klm{- \frac12\ \vec{\xi}_i^{\ T} C^{-1} \vec{\xi}_i}
\ee
The class of correlations we have considered is described by
\be
c_{\mu \nu} = \cases{R &for $\mu=m\sigma - \tau ,\
                                \nu=m\sigma - \epsilon,\
                                \sigma = 1, \ldots , p/m$ ,\\
                       &with  $\tau ,\epsilon = 0, \ldots , \klm{m-1} \quad 
                                \klm{\tau \neq \epsilon}$ \\
                      1 &for $\mu=\nu$ \\
                      0 & else \\}
\ee
The average over the fractional phase space of couplings is now performed with
\eref{avergauss} and results for $m \rightarrow \infty$ and 
$N \rightarrow \infty$ in
\bea
\label{entropym}
\fl
\frac1N \skl{\skl{ \rm{ln} V }}_{\xi} = 
\mathop{{\rm Extr}}_{q}  \Biggl\{
                & \ \alpha\  \mathop{{\rm Extr}}_{r,\hat r} \gkl{ 
                - i r \hat r - \frac12 c\ \klm{1-q}\ r^2 
                + \int Dt\ {\rm ln} H\klm{\omega} } \Biggr. \nonumber \\
                & \left. +\  \frac12\ {\rm ln} \klm{1-q}\
                +\ \frac12 \frac{q}{1-q} \right\}
\eea
with
\be
\omega = \frac{\kappa + \hat r + \sqrt{q}\ t}{\sqrt{1-q}}  \qquad ,
\qquad H\klm{x} = \int_x^{\infty} Dt
\ee
and $c = R m$ in the limit $m \rightarrow \infty$, 
so $-1<c\leq \infty$ since $-\klm{m-1}^{-1} < R \leq 1$. We have made the replica
symmetric ansatz for $q_{\alpha \beta}$ and the additional
order parameter $r_{\alpha} = m^{-1} \sum_{\mu = 1}^m x_{\alpha}^{\mu}$,
where $x_{\alpha}^{\mu}$ are the conjugate variables to the local fields
$\lambda_{\alpha}^{\mu}$, and its conjugate $\hat r_{\alpha}$. 

If we solve \eref{entropym} for the extremum we first can write $r$ in terms
of $\hat r$ and find in the limit $q \rightarrow 1$
\bea
\label{rhat}
- \frac{\hat r}c &=& \klm{\kappa +\hat r} H\klm{-\kappa - \hat r}
                  \ +\ \frac1{\sqrt{2 \pi}}
                    {\exp\gkl{- \frac12 \klm{\kappa +\hat r}^2}}
           \\
\label{alphahat}           
\alpha_c^{-1}\klm{c,\kappa} &=&  H\klm{-\kappa - \hat r}\ + \
          \kappa\ \frac{\hat r}c
\eea
From the first equation we can obtain $\hat r$ numerically and plug it into
the second to find $\alpha_c$. \Fref{alpham} shows the storage capacity as
a function of $c$ for several values of $\kappa$. For $\kappa = 0$, $\alpha_c$
approaches 1 as $c \rightarrow -1$. This behaviour is in agreement with                   
\eref{specialR} for large $m$. Fontanari and Meir (1989) considered the case
$m = p$, so all patterns are pairwise correlated. However, due to an error 
in one of the saddle point equations 
$\alpha_c\klm{\kappa=0}$ becomes less than 1 for values below $c \approx
- 0.7$ and reaches $\alpha_c = 0$ for $c \rightarrow -1$. 

For hierachically correlated biased patterns the storage capacity has
been calculated by Engel (1990). Above certain values of the bias the
critical capacity is less than two and for certain ranges even less than one.
It is not clear to us at which point general position is violated
in this case.

In summary, we have analyzed the behaviour of the storage capacity of a 
perceptron for correlated patterns. We find, that the storage capacity
is lowered with respect to uncorrelated patterns when different patterns
(negative overlap) are mapped onto the same output, but does not fall below
one, in agreement with Cover's theorem.
As a consequence we
suggest that the correlation matrix of the patterns should be analyzed
for problems which lead to a reduced storage capacity as for example the 
bit sequence. 

Future work should include the calculation of the stability of the replica
symmetric solution and the storage capacity for a binary perceptron.
One could also investigate the consequences for other architectures such as
the  commitee or parity machine, although we think that the results should
be similar for these cases.

After completion of this work we have learned that a similar problem
has been studied by Winkel (1995) using a different approach.

\ack
We would like to thank I Kanter for useful discussions.
This work has been supported by the Deutsche Forschungsgemeinschaft.

\References 
\item[] Bork A {\it Diploma Thesis}, Universit\"at W\"urzburg 1994
\item[] Cover T M 1965 {\it IEEE Trans. Electron. Comput.} {\bf EC--14} 326
\item[] Engel A 1990 \JPA {\bf 23} 2587
\item[] Fontanari J F and Meir R 1989 \JPA {\bf 22} L803       
\item[] Gardner E 1988 \JPA {\bf 21} 257
\item[] Gardner E and Derrida B 1988 \JPA {\bf 21} 271
\item[] Monasson R 1992 \JPA {\bf 25} 3701
\item[] Opper M 1988 \PR A {\bf38} 3824
\item[] Priel A, Blatt M, Grossmann T and Domany E 1994 \PR E {\bf 50} 577
\item[] Schr\"oder M, Eisenstein E, Kanter I and Kinzel W (unpublished) 
\item[] Winkel J {\it Diploma Thesis}, Universit\"at Regensburg 1995
\endrefs  

\Figures
\begin{figure}
\caption{The critical storage capacity $\alpha_c$ as a function of the
         overlap $R$ between two correlated patterns for $\kappa = 0$.
         The dots with their corresponding error bars are results from
         numerical simulations for systems with $N=100$.
         Inset: The median learning time to the power of $-1/2$ as a
         function of $\alpha$ for $R=-.95$. The values are averaged over
         1000 samples, the line is a least square fit for the data and
         the intersection between the extrapolation and the x-axis
         gives the estimated value of $\alpha_c$.}
\label{alpha2}
\end{figure}
\begin{figure}
\caption{The critical storage capacity $\alpha_c$ as a function of the
         correlation $c = R\ m$ for various values of 
         $\kappa= 0.0, 0.2, 1.0, 2.0$ (from top to bottom).}
         
\label{alpham}
\end{figure}
                                 
\include {pics}                         
        
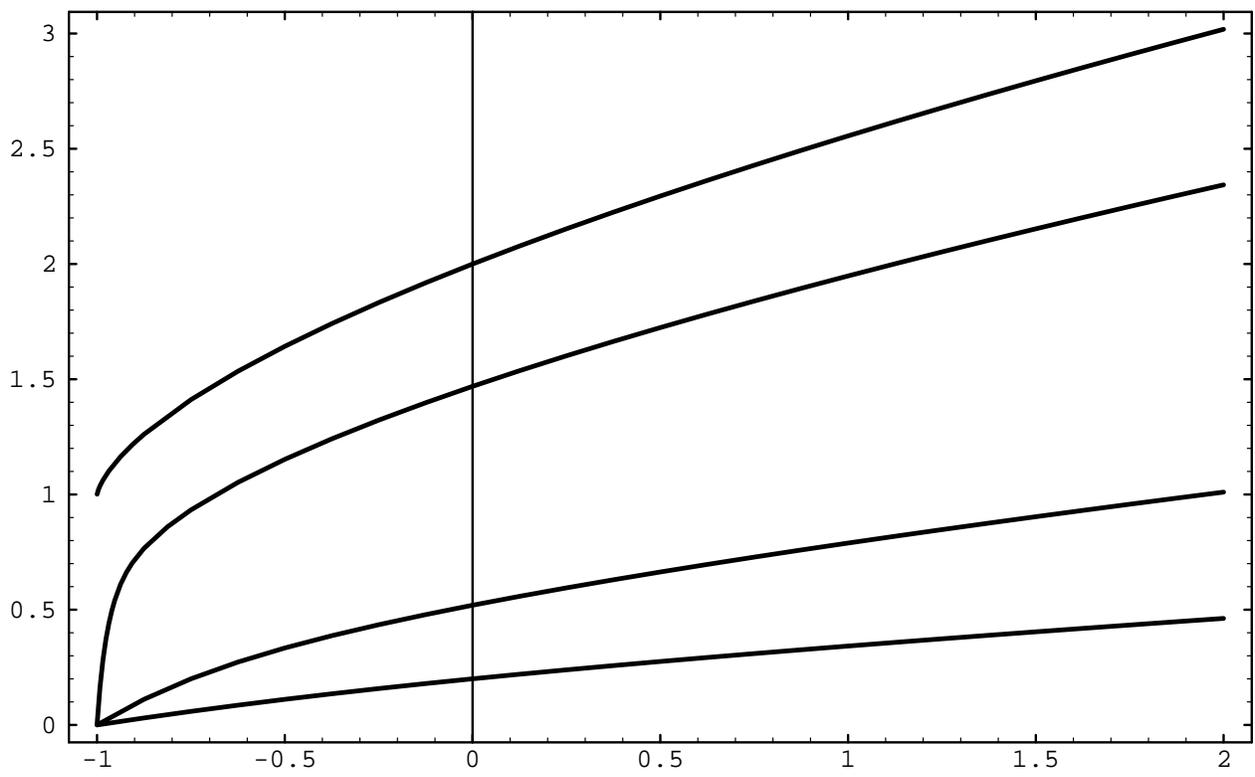
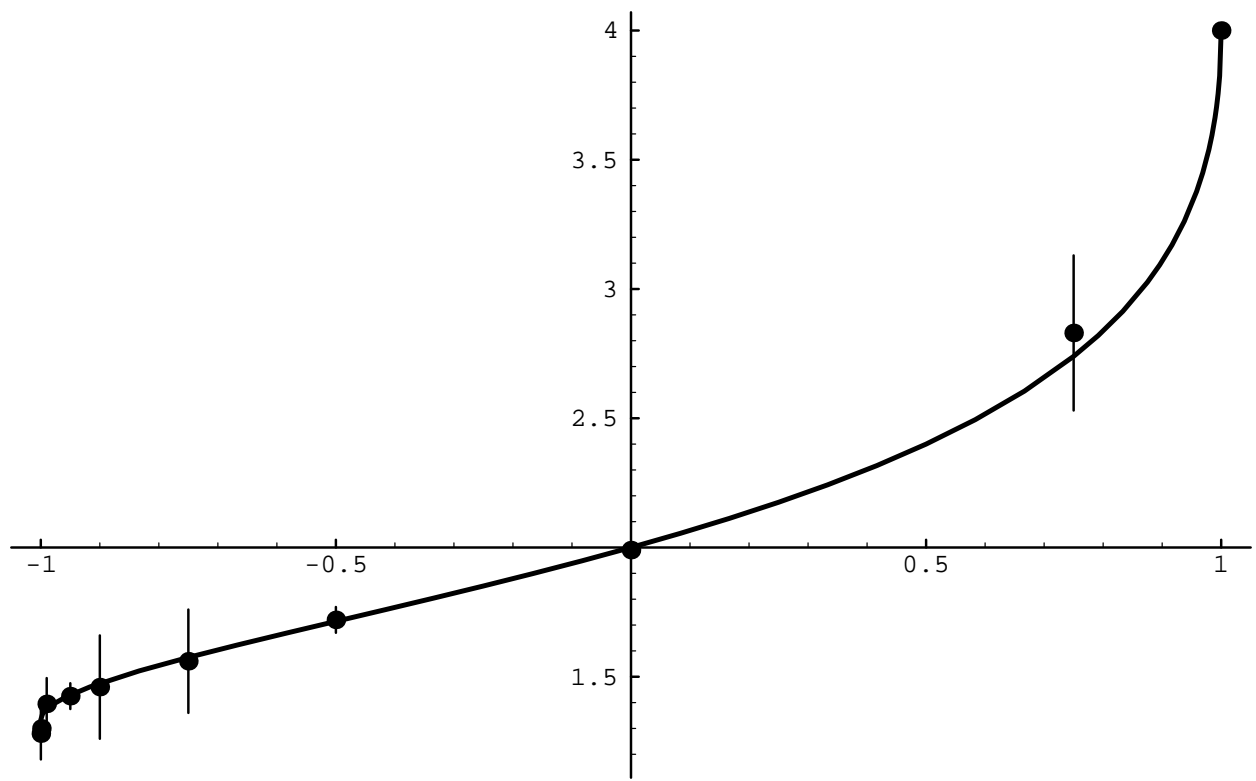
\end{document}